\documentclass[aps,notitlepage,superscriptaddress,nofootinbib]{revtex4-2} 
\usepackage[utf8]{inputenc}
\usepackage{bm}
\usepackage{bbm}
\usepackage{epsfig,wrapfig,fancybox}
\usepackage{float}
\usepackage{mathrsfs}
\usepackage{amsmath,amssymb,amsfonts}
\usepackage{graphicx,xcolor}

\usepackage[bookmarks=false, pdfstartview=FitH]{hyperref}

\newcommand{\continuousp}[2]{\ensuremath{p_{#1}(#2)}}
\newcommand{\discretep}[2]{\ensuremath{P_{\{#1\}}(#2)}}
\newcommand{\unitr}[1]{\ensuremath{R_{0,{#1}}}}
\newcommand{\innerr}[1]{\ensuremath{R_{i,{#1}}}}

\begin{document}

\title{Origin of Symmetry Breaking in the Grasshopper Model}

\author{David Llamas}
\affiliation{Department of Physics, University of Massachusetts Boston, Boston, MA 02125, USA}
\author{Jaron Kent-Dobias}
\email{jaron.kent-dobias@roma1.infn.it}
\affiliation{Istituto Nazionale di Fisica Nucleare, Sezione di Roma I, Rome, Italy}
\author{Kun Chen}
\affiliation{Center for Computational Quantum Physics, Flatiron Institute, 162 5th Avenue, New York, NY 10010, USA}
\author{Adrian Kent}
\email{apak@cam.ac.uk}
\affiliation{Centre for Mathematical Sciences, University of Cambridge, Wilberforce Road, Cambridge, CB3 0WA, United Kingdom.}
\affiliation{Perimeter Institute for Theoretical Physics, 31 Caroline Street North, Waterloo, Ontario N2L 2Y5, Canada.}
\author{Olga Goulko}
\email{olga.goulko@umb.edu}
\affiliation{Department of Physics, University of Massachusetts Boston, Boston, MA 02125, USA}
\date{\today}

\begin{abstract}
    The planar grasshopper problem, originally introduced in (Goulko \& Kent 2017 Proc. R. Soc. A 473, 20170494), is a striking example of a model with long-range isotropic interactions whose ground states break rotational symmetry. In this work we analyze and explain the nature of this symmetry breaking with emphasis on the importance of dimensionality. Interestingly, rotational symmetry is recovered in three dimensions for small jumps, which correspond to the non-isotropic cogwheel regime of the two-dimensional problem. We discuss simplified models that reproduce the symmetry properties of the original system in $N$ dimensions. For the full grasshopper model in two dimensions we obtain quantitative predictions for optimal perturbations of the disk. Our analytical results are confirmed by numerical simulations.
\end{abstract}

\maketitle

\section{Introduction}
Ref.~\cite{goulko2017grasshopper} introduced the following problem in geometric combinatorics: A grasshopper lands at a random point on a planar lawn of area one. It then jumps once, a fixed distance $d$, in a random direction. What shape should the lawn be to maximise the chance that the grasshopper remains on the lawn after jumping?  

The original motivation \cite{kent2014bloch} to study the grasshopper problem was to formulate and analyze new Bell inequalities for the
setup where two parties carry out spin measurements about randomly chosen axes separated by a fixed angle and obtain the spin correlations. For this purpose Ref. \cite{kent2014bloch} formulated the problem on the surface of a sphere, where the grasshopper's jump corresponds to the separation of the measurement axes on the Bloch sphere, and presented some partial results.   More detailed analytical results for the grasshopper problem on the surface of the sphere were presented in Ref.~\cite{chistikov2020globehopping}.  Ref.~\cite{chistikov2020globehopping} also analysed the problem on the one dimensional circle, which characterises a physically significant subset of qubit measurements often considered in quantum cryptography and quantum information theory.

Here we discuss the planar grasshopper problem and its analogues in higher dimensions. One surprising discovery from Ref.~\cite{goulko2017grasshopper} is that, while the jump setup is rotationally symmetric, the resulting optimal lawn shapes are not rotationally symmetric in two dimensions. This is true for arbitrary small jumps $d>0$. The specific shapes of the optimal lawns were computed numerically for a range of $d$. For small jumps, $d\lesssim \unitr{2}$, where $\unitr{2}=1/\sqrt{\pi}$ is the radius of the two-dimensional ball with unit volume, i.e.\ a unit area disk, the optimal lawns shapes resemble cogwheels. The distance between the cogs is approximately the jump length $d$. The periodic structure of the lawn boundary implies that the rotational symmetry of the jump setup reduces to a dihedral symmetry in two dimensions. While numerical results for the optimal lawn shapes and an analytical proof that the disk shaped lawn is never optimal have already been obtained, so far a fundamental understanding of this symmetry breaking was lacking. The goal of this paper is to systematically study the nature of the emergent symmetry breaking and its implications on the optimal grasshopper lawn shapes. As we will see, the symmetries of the solutions depend strongly on the dimensionality of the problem.

\section{Problem and Methods}

For a formal statement of the grasshopper problem in $N$ dimensions, we denote the lawn domain as $\Omega\subset{\mathbb R}^N$ and define the lawn function $\mu(\mathbf{r}):{\mathbb R}^N\to\{0, 1\}$ as the indicator function of $\Omega$, i.e., such that $\mu(\mathbf{r}) =1$ if and only if $\mathbf{r} \in \Omega$. Then
\begin{equation}
\int_\Omega dV=\int_{{\mathbb R}^N}d^Nr\,\mu  (\mathbf{r}) = 1 \, .
\label{eq:munorm}
\end{equation}
Note that in the most general version of the problem, lawns may have continuous density in the range $ ( 0, 1 ] $. The known numerical and theoretical results are consistent with there always being an optimal lawn with density $1$. The version of the problem restricting to density $1$ lawns is also natural, and we consider it here.

The grasshopper probability functional, $\continuousp{\mu}{d}$, is then given by 
\begin{equation}\label{successprob}
  \continuousp{\mu}{d} = \frac{1}{ S(N,d)} 
\int_{\Omega}dV_1 \int_{\Omega}dV_2\,\delta(\|\mathbf{r}_1-\mathbf{r}_2\|-d)= \frac{1}{ S(N,d)} 
\int_{{\mathbb R}^N}d^Nr_1 \int_{{\mathbb R}^N}d^Nr_2\,\mu(\mathbf{r}_1) \mu( \mathbf{r}_2 )  \delta(\|\mathbf{r}_1-\mathbf{r}_2\|-d),
\end{equation} 
where $S(N,d)$ is the surface area of the $N$-dimensional sphere with radius
$d$, i.e., the area of the region accessible to one jump. Specifically, in two
dimensions, the surface area (circumference) of the two-dimensional sphere
(disk) equals $S(2,d)=2\pi d$, and in three dimensions, the corresponding area equals
$S(3,d)=4\pi d^2$.

The grasshopper probability \eqref{successprob} can be evaluated exactly when the lawn shape is an $N$-ball with radius $R_{0,N}$, by performing the integral \eqref{successprob} explicitly in generalized spherical coordinates. For the special case of a 2-ball, i.e.\ disk, the probability equals
\begin{equation}
\continuousp{\rm disc}{d}=1-\frac{2}{\pi}\left(\frac{d}{2\unitr{2}}\sqrt{1-\left(\frac{d}{2\unitr{2}}\right)^2}+\arcsin\left(\frac{d}{2\unitr{2}}\right)\right)
\label{eq:unitdiscprob}
\end{equation}
assuming $d\leq\unitr{2}$. The success probability approaches one as $d\rightarrow0$ and decreases monotonically with increasing $d$. In three dimensions, the corresponding probability for jumps smaller than the 3-ball radius, $d\leq\unitr{3}= (3/(4\pi))^{1/3}$, can be computed as
\begin{equation}
  p_{\rm ball}(d) = 1 - \frac{3}{4}\frac{d}{\unitr{3}} + \frac{1}{16} \left(\frac{d}{\unitr{3}}\right)^3
    \label{eq:unitballprob}
\end{equation}

Working with the double area integral of a $\delta$-distribution is difficult in more complicated geometries. 
To overcome this, we exploit the insight that the general probability functional from Eq.~\eqref{successprob} can be recast as a boundary rather than a volume integral, following a procedure previously used in the study of magnetic fluid monolayers \cite{McConnell_1988_Shapes, McConnell_1992_Note, kentdobias2015dipoles}. Specifically,
\begin{equation}
    p_\mu(d)=\int_{\Omega}dV_1\int_{\Omega}dV_2\,f(\|\mathbf{r_1}-\mathbf{r_2}\|)=-\int_{\partial\Omega}dS_1\int_{\partial\Omega}dS_2\,\Phi(\|\mathbf{r_1}-\mathbf{r_2}\|)(\hat{\mathbf{n}}_1\cdot\hat{\mathbf{n}}_2),
    \label{eq:surfaceformulation}
\end{equation}
where $\partial\Omega$ denotes the boundary of the lawn $\Omega$. Here we abbreviate the interaction as
\begin{equation}
    f(r)=\frac{1}{S(N,d)}\delta(r-d) 
\end{equation}
and define $\Phi$ to be a radially symmetric solution to $\nabla^2\Phi(r)=f(r)$ (in spherical coordinates). The vectors $\hat{\mathbf{n}}$ are the normal unit vectors orthogonal to the lawn boundary and $dS$ denotes the surface area element corresponding to the volume element $dV$.

To prove Eq.~\eqref{eq:surfaceformulation} we use the divergence theorem (Gauss's theorem).\footnote{In Ref.~\cite{kentdobias2015dipoles} the analogous derivation was performed using Green's theorem, which is equivalent to the current proof in two dimensions.} We also use the fact that $\nabla_1\Phi(\|\mathbf{r}_1-\mathbf{r}_2\|)=-\nabla_2\Phi(\|\mathbf{r}_1-\mathbf{r}_2\|)$. We write
\begin{eqnarray}
    p_\mu (d)
    &=&\int_{\Omega}dV_1\int_{\Omega}dV_2\,\nabla_1^2\Phi(\|\mathbf{r}_1-\mathbf{r}_2\|)
    =\int_{\Omega}dV_1\int_{\Omega}dV_2\,\nabla_1\cdot\nabla_1\Phi(\|\mathbf{r}_1-\mathbf{r}_2\|) \notag \\
    &=&\int_{\Omega}dV_2\int_{\partial\Omega}d\mathbf{S}_1\cdot\nabla_1\Phi(\|\mathbf{r}_1-\mathbf{r}_2\|)
    =-\int_{\Omega}dV_2\int_{\partial\Omega}d\mathbf{S}_1\cdot\nabla_2\Phi(\|\mathbf{r}_1-\mathbf{r}_2\|)\\
    &=&-\int_{\partial\Omega}\int_{\partial\Omega}d\mathbf{S}_1\cdot d\mathbf{S}_2\,\Phi(\|\mathbf{r}_1-\mathbf{r}_2\|) \, . \notag
\end{eqnarray}

In the last step we used the gradient version of the divergence theorem.
Note that $d\mathbf{S}=\hat{\mathbf{n}}\,dS$, matching \eqref{eq:surfaceformulation}.

We can work out explicitly the form of the function $\Phi$ in the surface
integrals. The Laplacian equation for radially symmetric functions in $N$
dimensions is
\begin{equation}
  \frac1{S(N,d)}\delta(r-d)=f(r)=\nabla^2\Phi(r)=\frac1{r^{N-1}}\frac\partial{\partial r}\left(r^{N-1}\frac{\partial \Phi}{\partial r}\right) \, . 
\end{equation}
This can be solved by explicit integration.
For 2D, the result is
\begin{equation}
  \Phi(r)=\frac1{2\pi}\Theta(r-d)\log\frac rd  \, ,
\end{equation}
while for all $N>2$ the result is
\begin{equation}
  \Phi(r)=\frac d{(N-2)S(N,d)}\left[\Theta(r-d)\left(1-\frac{d^{N-2}}{r^{N-2}}\right) - 1\right],
\end{equation}
where $\Theta$ is the Heaviside function.
$\Phi$ is not unique, since one can add to it any function with zero Laplacian
and still satisfy its definition. For $N>2$ we have chosen a convention where
$\Phi(r)$ approaches zero at large $r$, which will be useful in the next
section when we want to neglect the contribution of distant boundaries.

\section{A minimal model: the infinite half-space}
\label{sec:minmodel}

The goal of this section is to analyze the stability of the isotropic lawn in the limit of small $d$ in a way that generalizes to arbitrary dimension $N$. We will see that in two dimensions there is the potential for instability to small perturbations, as found for the disk in \cite{goulko2017grasshopper}, while three and higher dimensions such instabilities do not exist.

The starting point is the isotropic solution.  
Intuitively, one could think that in the limit $d\ll1$ this can be approximated as having zero curvature, i.e. as a half-plane (2d) or half-volume (3d).  One needs to be careful in basing arguments on this intuition, since the problem is defined globally rather than locally, and since optimal lawns for small $d$ need not necessarily
have small curvature.   Also, since the half-spaces have infinite volume, the lawn volume is not normalizable, and 
(\ref{successprob}) is not well-defined.   Nonetheless, as we will explain,
a perturbative analysis of the half-space problem in $N$ dimensions gives very useful insights into the standard problem framed above.

One could define a version of the problem on the half-space by assuming that the perturbed lawn boundary is a periodic function of a lattice, then normalizing by restricting to one lattice cell and to jumps from initial points within some given distance $D$ from the unperturbed boundary, then taking the limit of large $D$.  
Here we simplify the discussion by considering the contribution of small plane wave perturbations, which are 
periodic in $1$ direction rather than $N-1$ independent directions (i.e. are the general periodic
perturbations only for $N=2$).    We calculate their differential contribution to 
an expression of the form (\ref{successprob}) for the half-space, which is a meaningful
expression assuming a normalization as described.

Let $\mathbf x\in\mathbb R^{N-1}$ parameterize the boundary of the lawn and
$\mathbf r\in\mathbb R^N$. We consider a boundary defined by $\mathbf r(\mathbf
x)=(\mathbf x,\epsilon h(\mathbf x))$, where the height function $h$ gives a
smooth perturbation from the isotropic solution and its magnitude $\epsilon$ is such that $\epsilon\ll d$. The unnormalized normal vector to this surface is $\mathbf n(\mathbf x)=\nabla_\mathbf
r(r_N-\epsilon h(\mathbf x))=(-\epsilon\nabla_\mathbf xh(\mathbf x),1)$, where $r_N$ is the $N$th component of $\mathbf{r}$. Then,
the unnormalized probability of remaining on the lawn is given by the double surface integral
\begin{equation}
  p
  =-\int_{\mathbb R^{N-1}}d\mathbf x\,d\mathbf y\,\mathbf n(\mathbf x)\cdot\mathbf n(\mathbf y)\,\Phi(\|\mathbf r(\mathbf x)-\mathbf r(\mathbf y)\|)
  =-\int_{\mathbb R^{N-1}}d\mathbf x\,d\mathbf y\,\big(1+\epsilon^2\nabla h(\mathbf x)\cdot\nabla h(\mathbf y)\big)\,\Phi(\|\mathbf r(\mathbf x)-\mathbf r(\mathbf y)\|) \, .
\end{equation}
The stability of the flat plane to small perturbations is governed by the coefficient for $p$ at second order in $\epsilon$. Expanding $p-p_0$ in $\epsilon$, we find
\begin{equation} \label{eq:plane.expansion}
  p-p_0
  =-\epsilon^2\int_{\mathbb R^{N-1}} d\mathbf x\,d\mathbf y\,\left[
    \nabla h(\mathbf x)\cdot\nabla h(\mathbf y)\,\Phi(\|\mathbf x-\mathbf y\|)
    +\frac12\frac{(h(\mathbf x)-h(\mathbf y))^2}{\|\mathbf x-\mathbf y\|}\Phi'(\|\mathbf x-\mathbf y\|)
  \right]
  +O(\epsilon^4) \, .
\end{equation}
To examine the effect of plane-wave perturbations, we take $h(\mathbf x)=\cos(kx_1)$.
We expect the effect of such a perturbation to be proportional to the volume of the surface,
so we define the stability coefficient $\delta p_k(d)$ to be
\begin{equation}
  p-p_0=\delta p_k(d)\epsilon^2L^{N-1}+O(\epsilon^4),
\end{equation}
where $L$ is the linear dimension of the problem, which we will eventually take to infinity.
Details of the calculation of this coefficient can be found in Appendix~\ref{sec:half-space.details}. The result is
\begin{equation} \label{eq:plane.instability}
  \delta p_k(d)
  =\frac{\Gamma(\frac N2)}{2\sqrt\pi d}\bigg[
    \frac{J_{\frac{N-3}2}(kd)}{(kd/2)^{(N-3)/2}}
    -\frac1{\Gamma(\frac{N-1}2)}
  \bigg],
\end{equation}
where $J_m$ is the $m$th Bessel function. For $N=2$ and $N=3$, we have
\begin{align}\label{twothreedperts}
  \delta p_k^{(N=2)}(d)=\frac1{2\pi d}[\cos(kd)-1],
  &&
  \delta p_k^{(N=3)}(d)=\frac1{4d}[J_0(kd)-1] \, ,
\end{align}
respectively.

\begin{figure}
  \centering
  \includegraphics{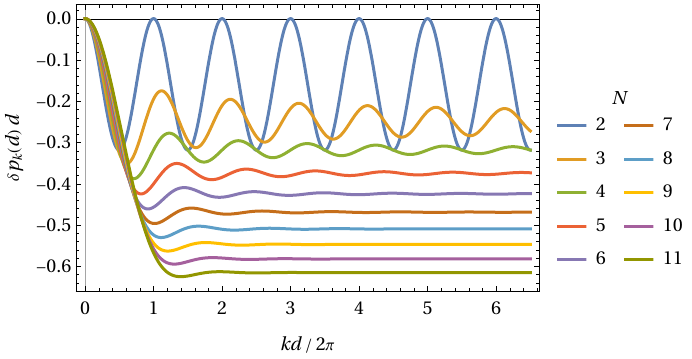}
  \caption{
    The stability of flat half-space in $N$ dimensions to plane-wave
    perturbations of wavenumber $k$. All values are negative except at $k=0$
    (which corresponds to translation of the interface and is not a
    probability-conserving perturbation) and for $kd$ that are multiples of
    $2\pi$ in the case $N=2$. These values are zero, meaning that there is a
    marginal stability to perturbations with wavenumber commensurate with $d$
    in two dimensions but not an instability. For larger dimension $N$ of
    space, the result becomes increasingly insensitive to the commensurability
    of $k$ and $d$.
  } \label{fig:plane_stability}
\end{figure}

Some of the resulting functions can be seen plotted in
Fig.~\ref{fig:plane_stability}. First, notice that for $N\geq3$ and $k\neq0$,
the result is always negative. This means that for three and higher dimensions the lawn is
stable to plane wave instabilities. For $N=2$, there is a point of marginal
stability for each integer multiple of $kd/2\pi$, which corresponds to
wavenumbers that are commensurate with the jump size $d$. As $N$ is increased,
the dependence on commensurability with $d$ becomes steadily weaker. There is a
simple intuition for this.  The average absolute value of the overlap of two random unit vectors in $N$ dimensions is $\frac{1}{\sqrt{N}}$.   Hence the expected projection of the
jump onto the wavevector of a plane wave instability shrinks like $dN^{-1/2}$.
The diminishing probability of alignment between the jump and the wavevector
pushes any oscillations to higher wavevectors and the variation of this
alignment causes the oscillations to decay. Indeed, the curves for various $N$ collapse like $N^{-1/2}d\,\delta p_k(N^{1/2} x/k)$ with $x=N^{-1/2}kd$, approaching a limit curve of the form
\begin{equation}
  \lim_{N\to\infty}N^{-1/2}d\,\delta p_k(N^{1/2}x/k)=\frac1{2\sqrt{2\pi}}(e^{-x^2/2}-1),
\end{equation}
with oscillations totally suppressed.

Before delving into the instabilities inherent to the two-dimensional disk,
it is worth commenting on the potential role of higher-order perturbations,
i.e., $\epsilon^4$ and greater, in determining the fate of marginal stability
for $N=2$. One might argue that a complete picture would require incorporating
these higher-order terms. However, our study focuses on a finite jump size $d$,
and as we will show, this alone is sufficient to drive a compact system from a
marginally stable state to true instability. Thus, while high-order perturbations may offer nuanced insights into stability in the half-space version of the problem, they are not relevant to our present investigation of the
stability of compact regions in the plane, which centers on the transition driven by finite-size effects.

\section{Instabilities of the two-dimensional disk}

Even with the simplification of moving from the area integrals to line
integrals, there are few two dimensional geometries where the problem is analytically
tractable. Besides the half-space analyzed above, the disk can also be treated
explicitly, as well as small perturbations around the disk. Here, we will
compute the second variation of the grasshopper probability for the disk with
respect to small perturbations of various discrete symmetry breaking orders,
explicitly confirming the findings of \cite{goulko2017grasshopper}.

The information provided by this calculation will \emph{not} tell when a cog of
a given number of teeth will be the most stable configuration, since the cog is
a large perturbation to the circle. Instead, this variation of the probability describes a kind of
spinodal: when does an infinitesimal deformation of the disk towards a given
cog raise the probability instead of lowering it? These spinodal points
therefore provide some information about when symmetry breaking of a given kind
becomes advantageous to isotropy.

Consider a two-dimensional lawn whose boundary is given parametrically in standard polar
coordinates by
\begin{equation}\label{repsilon}
  r_{n,\epsilon}(\theta)=R_\epsilon+\epsilon\cos(n\theta ) \, .
\end{equation}
The lawn function, also in polar coordinates, is
\begin{equation}
  \mu_{n,\epsilon}(\mathbf r)=
  \begin{cases}
    1&r<r_{n,\epsilon}(\theta) \, , \\
    0&\text{otherwise} \, .
  \end{cases}
\end{equation}
Here, the radius $R_\epsilon$ is picked to preserve the area of the lawn. We
are interested in small perturbations, and in particular in the stability of
the probability  of the disk. Specifically, we will evaluate
\begin{equation}
  p_{n,\epsilon}(d)=p_\mathrm{disc}(d)+\delta p_n(d)\epsilon^2+O(\epsilon^4) \, .
\end{equation}
When the coefficient $\delta p_n(d)$ is negative, the perturbation lowers the
probability and the circular lawn is stable to sinusoidal perturbations of $n$-fold
symmetry. When it is positive, the probability is increased by the perturbation
and the circular lawn is unstable.

Because we only evaluated the quadratic coefficient of the probability in
$\epsilon$, we only need the area $R_\epsilon$ to quadratic order. Since normalization requires that
\begin{equation}
  1
  =A
  =\int_{\mathbb R^2}d\mathbf r\,\mu_{n,\epsilon}(\mathbf r)
  =\int_0^{2\pi}d\theta\int_0^{r_{n,\epsilon}(\theta)}dr\,r
  =\frac12\int_0^{2\pi}d\theta\,\big(R_\epsilon+\epsilon\cos(n\theta)\big)^2
  =\pi R_\epsilon^2+\frac12\pi\epsilon^2 \, ,
\end{equation}
we must set
\begin{equation}
  R_\epsilon
  =\sqrt{R_{0,2}^2-\frac12\epsilon^2}
  =R_{0,2}-\frac1{4R_{0,2}}\epsilon^2+O(\epsilon^4)
\end{equation}
where $R_{0,2}=1/\sqrt\pi$. To treat the line integral, we change the variable of integration from the line
element $dS$ to the angular one $d\theta$. This produces a correction to the
measures given by the norm of the tangent to the curve, or
\begin{equation}
  \frac{dS}{d\theta}=\|\mathbf r'_{n,\epsilon}(\theta)\| \, .
\end{equation}
where the prime denotes the derivative of a function with respect to its parenthetical argument.
Noticing that $\hat{\mathbf n}_1\cdot \hat{\mathbf n}_2=\hat{\mathbf
t}_1\cdot\hat{\mathbf t}_2$ (the scalar product of unit normals is the same as
the scalar product of unit tangents for a 1D curve), and that $\hat{\mathbf
t}=\mathbf r'(\theta)/\|\mathbf r'(\theta)\|$, we find
\begin{equation}
  \begin{aligned}
    p_{n,\epsilon}(d)
    &=-\int_0^{2\pi}d\theta_1\,\|\mathbf r'_{n,\epsilon}(\theta_1)\|
    \int_0^{2\pi}\,d\theta_2\,\|\mathbf r'_{n,\epsilon}(\theta_2)\|\,
    \Phi\big(\|\mathbf r_{n,\epsilon}(\theta_1)-\mathbf r_{n,\epsilon}(\theta_2)\|\big)
    \,\hat{\mathbf n}(\theta_1)\cdot\hat{\mathbf n}(\theta_2) \\
    &=-\int_0^{2\pi}d\theta_1\int_0^{2\pi}d\theta_2\,
    \Phi\big(\|\mathbf r_{n,\epsilon}(\theta_1)-\mathbf r_{n,\epsilon}(\theta_2)\|\big)\;
    \mathbf r_{n,\epsilon}'(\theta_1)\cdot\mathbf r_{n,\epsilon}'(\theta_2),
  \end{aligned}
\end{equation}
where the scalar product is now between the \emph{unnormalized} tangent
vectors. Finally, we change integration variables to $\theta=\theta_1$,
$\phi=\theta_2-\theta_1$, and have
\begin{equation} \label{eq:disk.integrals}
  p_{n,\epsilon}(d)
  =-\int_0^{2\pi}d\theta\int_{-\pi}^{\pi}d\phi\,
  \Phi\big(\|\mathbf r_{n,\epsilon}(\theta)-\mathbf r_{n,\epsilon}(\theta+\phi)\|\big)\;
  \mathbf r_{n,\epsilon}'(\theta)\cdot\mathbf r_{n,\epsilon}'(\theta+\phi),
\end{equation}
a convenient expression because the integrand in the limit $\epsilon=0$ is everywhere symmetric in
$\phi\mapsto-\phi$.

\begin{figure}
  \includegraphics{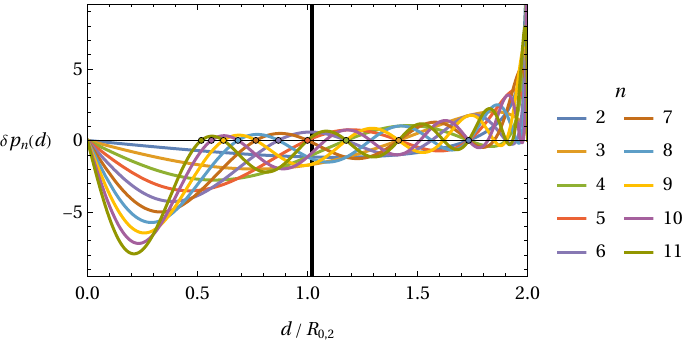}

  \caption{
    The coefficient of stability for the disk to small perturbations of
    $n$-fold symmetry. A large dot is drawn at the smallest value of $d$ where
    each curve first becomes positive. The black bar shows the point in $d$ at
    which the transition to disconnected shapes occurs as measured in
    \cite{goulko2017grasshopper}.
  } \label{fig:disk_stability} 
\end{figure}

\begin{figure}
  \includegraphics{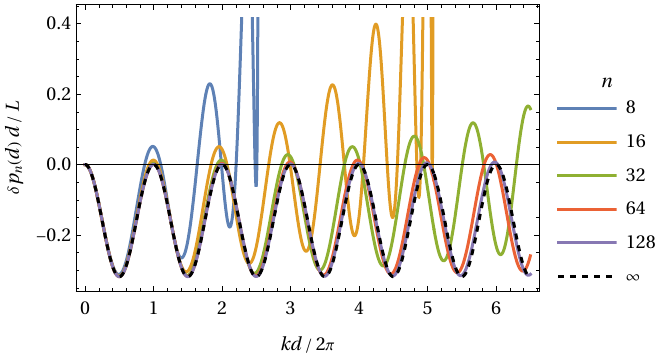}

  \caption{
    The scaled coefficient $\delta p_n$ for increasing $n$ on the disk as a function of $kd=\sqrt\pi nd$, along
    with the result for the semi-infinite plane ($n=\infty$) from \eqref{eq:plane.instability}. As $n$ is
    increased, the disk result asymptotically approaches that of the
    semi-infinite plane. Since the finite-$n$ curves tend to the zeroes of the limit curve from above,
    the marginally stable points of the half-plane are destabilized at finite
    $n$.
  } \label{fig:convergence}
\end{figure}

The detailed expansion of \eqref{eq:disk.integrals} to second order in
$\epsilon$ and the evaluation of the resulting integrals can be found in
Appendix~\ref{sec:disk.details}. The resulting stability coefficient is
\begin{equation} \label{eq:disk.delta.prob}
  \delta p_n(d)=-\frac{\cos\phi_0-\cos(n\phi_0)}{\sin\phi_0},
\end{equation}
where $\phi_0$ is given by
\begin{equation}
  \phi_0=\cos^{-1}\left(1-\frac{d^2}{2R_{0,2}^2}\right) 
\end{equation}
for $0 \leq d \leq \frac{2}{\sqrt{\pi}}=2\unitr{2}$. 
This corresponds to the angle at which $\|\mathbf r_{n,0}(\theta)-\mathbf
r_{n,0}(\theta+\phi_0)\|=d$, i.e., the difference in angle along the circle at
which that distance between two points is equal to the jump size. This function
is plotted for several $n$ in Fig.~\ref{fig:disk_stability}. There, one can see
that the local maxima of each curve are positive, indicating instabilities in
their vicinity.

The limit of \eqref{eq:disk.delta.prob} as $n\to\infty$ asymptotically
approaches the result (\ref{twothreedperts}) found in the previous section for the infinite
half-space, as expected. The convergence can be seen in
Fig.~\ref{fig:convergence}. The mechanism by which the disk is unstable at
small $d$ for a large number of oscillations $n$ is revealed: the approach of
the finite-$n$ curves to the zeroes of the limit curve is from above. Therefore, the points
at which the wavevector $k$ is commensurate with the jump size $d$, which are
\emph{marginally stable} perturbations of the flat interface, become
\emph{unstable} in this case. The curves of other finite geometries might
approach the limit from the other direction, in which case they would be stable
against perturbations of this type.

The first zero of $\delta p_n(d)$ for fixed $n$ is significant because it indicates the
smallest jump size $d$ for which the disk is unstable to infinitesimal
perturbations of $n$-fold symmetry of the form (\ref{repsilon}).
The first zero of $\delta p_n(d)$ for given
$n$ corresponds to
\begin{equation}
  \phi_0=\frac{2\pi}{n+1}
\end{equation}
which corresponds to zeros at
\begin{equation}
  d_0=R_{0,2}\sqrt{2(1-\cos\phi_0)}=\sqrt{\frac2\pi\left(1-\cos\frac{2\pi}{n+1}\right)}
\end{equation}
whose large-$n$ behavior is $d_0\simeq2\sqrt\pi n^{-1}$, in agreement with the prediction from \cite{goulko2017grasshopper}.

%

The stability is examined by seeing where the function first becomes positive.
For instance, $\delta p_2(d)$ is negative for small $d$, and crosses zero at
$d=\sqrt{3/\pi}$, where the circular lawn becomes unstable to elongation. This
expression, along with all even-order perturbations, diverges at
$d=2R_{0,2}$, the diameter of the disk, while odd-order perturbations
approach zero. Intuitively, the even-order perturbations allow longer jumps
across the disk's diameter, while the odd ones do not.

\begin{figure}
  \includegraphics{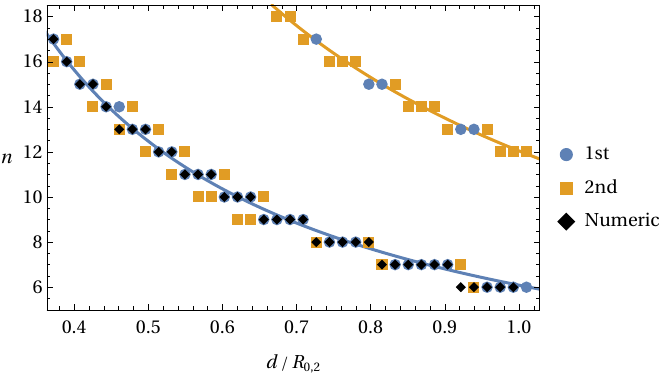}\hfill
  \caption{
    This plot shows the 1st and 2nd most unstable modes $n$ at a given $d$ for
    $n\leq18$. The blue and yellow lines show the location of the first and
    second peak in $\delta p_n(d)$, respectively.  The location of the first
    peak is extremely similar to (3.4) of \cite{goulko2017grasshopper} but not
    identical. If the cutoff in $n$ is increased, more bands appear at higher
    $n$. Black markers denote the corresponding numbers of cogs for the optimal solutions found through numerical simulation in Ref.~\cite{goulko2017grasshopper} (the same data are shown in the left panel of Fig.~5 in Ref.~\cite{goulko2017grasshopper}). The numerical results from \cite{goulko2017grasshopper} are in very close agreement with the current prediction.  
  } \label{fig:maximally_unstable_mode}
\end{figure}

\begin{figure}
  \includegraphics{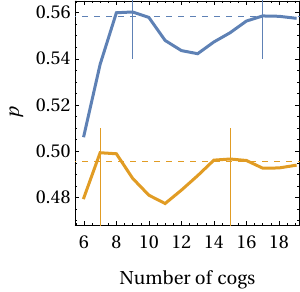}
  \hspace{1em}
  \includegraphics{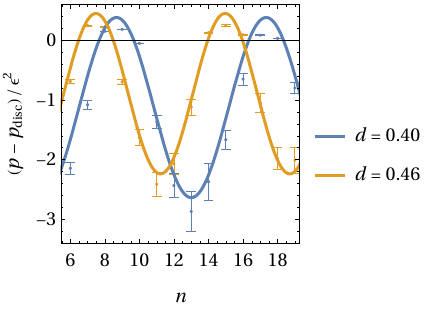}

  \caption{
    \textbf{Left:} Grasshopper probability for lawns with
    fixed cog number for two values of $d$, obtained numerically for discrete lawns with $10,000$ cells. The maxima corresponding to the two
    leading unstable modes are clearly visible ($n=9$ and $n=17$ for $d=0.4$;
    $n=7$ and $n=15$ for $d=0.46$) and are marked with thin vertical lines.
    These are indeed (local) maxima of the grasshopper problem. Horizontal
    dashed lines denote the corresponding disk probabilities given by
    Eq.~\eqref{eq:unitdiscprob}.
    \textbf{Right:} The same probability rescaled by the squared amplitude
    $\epsilon$ of the cogs (symbols with error bars). The solid lines show $\delta p_n(d)$ from
    \eqref{eq:disk.delta.prob}, which corresponds roughly with the
    finite-$\epsilon$ numerical data.
  } \label{fig:cog.number}
\end{figure}

In addition to the first moment of instability, we can analyze which modes are
maximally unstable for a given $d$, i.e., which modes maximize $\delta p_n(d)$.
For almost all $d$, any purported maximum at some value $n$ will be
improved by another much higher value. However, 
we expect that these high frequency modes will
be severely penalized by higher-order terms in the expansion of
$p_{n,\epsilon}(d)$, and so to find better candidates for the relevant
symmetry-breaking modes it is useful to truncate the range of $n$ considered.
In Fig.~\ref{fig:maximally_unstable_mode}, we show the values of the two most
unstable modes $n$ at a variety of fixed $d$ for $n$ truncated at 18. As
evidenced in that figure, the first or second most unstable mode corresponds
fairly closely with the actual ground-state symmetry found numerically in \cite{goulko2017grasshopper}. Both roughly follow a trend that is described by the location of
the first peak in $\delta p_n(d)$.

The stability coefficient \eqref{eq:disk.delta.prob} is somewhat predictive of
what shapes are favored at nonzero $\epsilon$. Fig.~\ref{fig:cog.number} shows
the numerical probabilities for lawns with some fixed number of cogs for two
different values of $d$. The peaks of these curves correspond well with those
predicted by the peaks in the stability coefficient. Moreover, when the
probability is rescaled by the amplitude of the cogs, the entire dependence of
its variation with $n$ is roughly predicted by our formula. Of course, they do
not correspond exactly, since with nonzero $\epsilon$ there are higher-order
corrections that must enter. In addition, we cannot predict the optimal
amplitude $\epsilon$ without going to higher order.

\section{Numerical results in 3d}
In previous work \cite{goulko2017grasshopper} numerical results were presented for the grasshopper problem on the 2d plane. In this section we present a similar discussion for three dimensions. The numerical setup in 2d involves defining a discrete version of the grasshopper problem by dividing the plane into grid cells and assigning a spin variable $s_i$ to the center of each cell $i$, where $s_i = 0$ and $s_i = 1$ correspond to the cell being unoccupied or occupied by the lawn, respectively. This setup easily generalizes to higher dimensions. The discrete version of the functional $\continuousp{\mu}{d}$ from Eq.~\eqref{successprob} becomes
\begin{equation}
  \discretep{s}{d} =\frac{1}{ S(N,d)}  \frac{1}{M^{2}h}\sum_{i, j}  s_i s_j \phi \left( \frac{\| \mathbf{r}_i -\mathbf{r}_j  \| - d}{h} \right) \, , \label{eq:spinham}
\end{equation}
where $M$ denotes the number of spins and $h$ the edge length of a unit cell. The normalization is $Mh^N=1$. The discretization involves a smoothed approximation $\delta(r)\rightarrow\delta_h(r)=\phi(r/h)/h$ to the (one-dimensional) $\delta$-function, which appears in (\ref{successprob}). As in \cite{goulko2017grasshopper}, we choose $\phi(r/h)=\left(1+\cos(\frac{\pi r}{2h})\right)/4$ if $\|r\|\leq 2h$ and 0 otherwise, following \cite{peskin2002deltafn}.

In the continuum limit, $\discretep{s}{d}\rightarrow \continuousp{\mu}{d}$. To resolve the jump distance $d$ we require $h \ll d$. We look for spin configurations that maximize $\discretep{s}{d}$, which represent solutions to the discrete grasshopper problem considered. This problem is equivalent to finding the ground state of a spin system with
Hamiltonian $H=-\discretep{s}{d}$. This spin system is a conserved spin Ising model with fixed-range interactions, where the interaction range $d$ is large. The optimal configurations are found using simulated annealing or parallel tempering algorithms \cite{goulko2017grasshopper}.

\begin{figure}[bt]
    \centering
    \includegraphics[clip=true,width=0.49\textwidth]{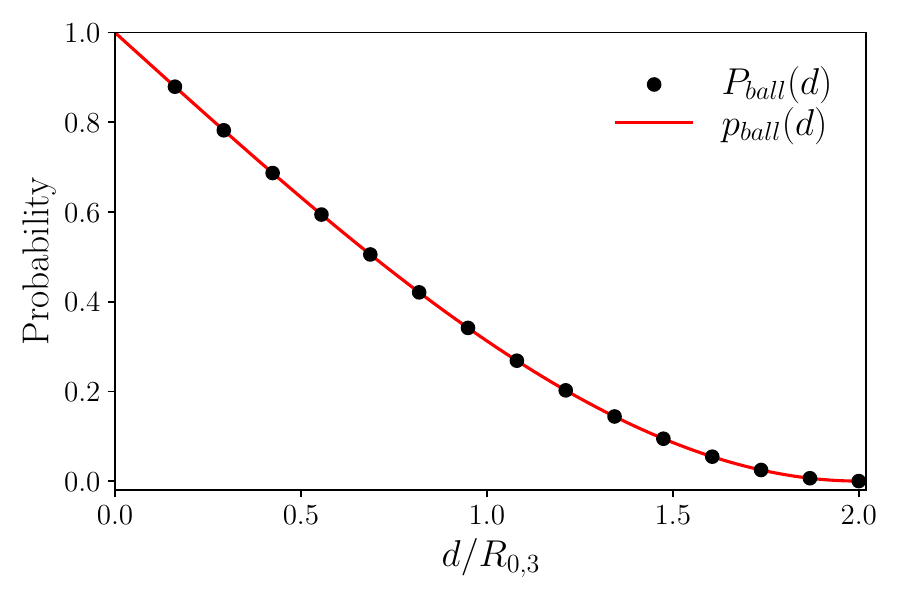}\hfill
\includegraphics[clip=true,width=0.49\textwidth]{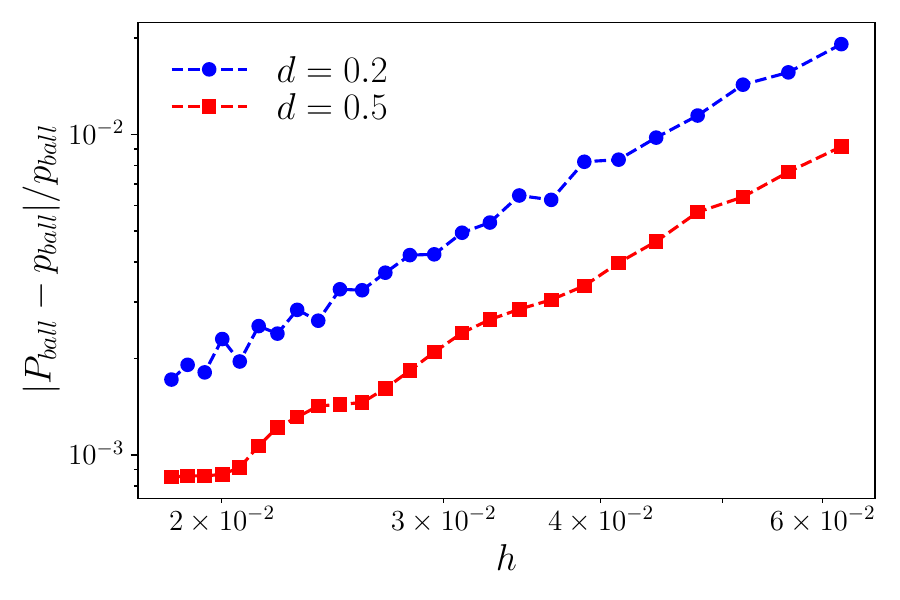}
    \caption{Study of discretization effects. Left panel: The exact continuous probability functional $\continuousp{\mu}{d}$ for the solid 3-ball of unit volume given by Eq.~\eqref{eq:unitballprob} (red solid line) compared with the corresponding discrete $\discretep{s}{d}$ (black dots) as function of the grasshopper jump distance $d$. For $d\leq\unitr{3}$ the 3-ball configuration is the optimal lawn shape.
    Right panel: Relative deviation of $\discretep{s}{d}$ for the solid 3-ball configuration from the corresponding $\continuousp{\mu}{d}$ as function of the lattice spacing $h$ for two representative values of the grasshopper jump: $d=0.2\approx0.3\unitr{3}$ (blue dots and line) and $d=0.5\approx0.8\unitr{3}$ (red dots and line). For the highest resolutions considered ($M\approx 113,000$) the discretization error is well below $0.3\%$. Lines are to guide the eye.
    \label{fig:ball_probabilities}}
\end{figure}

To quantify discretization effects in three dimensions, in Fig.~\ref{fig:ball_probabilities} we compare the probability for the discretized solid 3-ball to the exact continuous model probability~(\ref{eq:unitballprob}). An analogous analysis for two dimensions was presented in \cite{goulko2017grasshopper}. We can see that as in \cite{goulko2017grasshopper} the discrete model is an excellent approximation of the continuous grasshopper problem and the discretization errors are of the order of $0.3\%$ or less at the highest resolution considered.
\begin{figure}
    \centering
    \includegraphics[clip=true,width=0.49\textwidth]{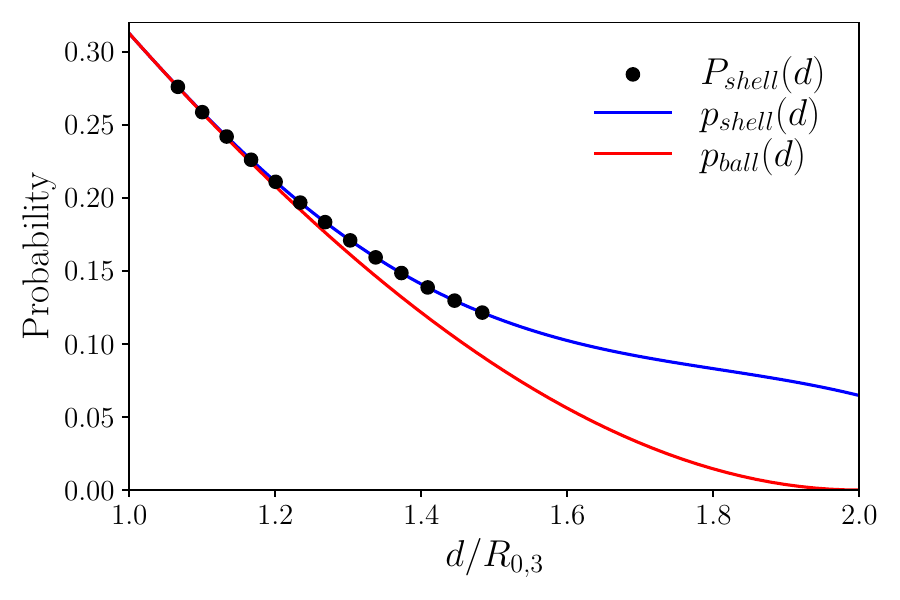}\hfill
\includegraphics[clip=true,width=0.49\textwidth]{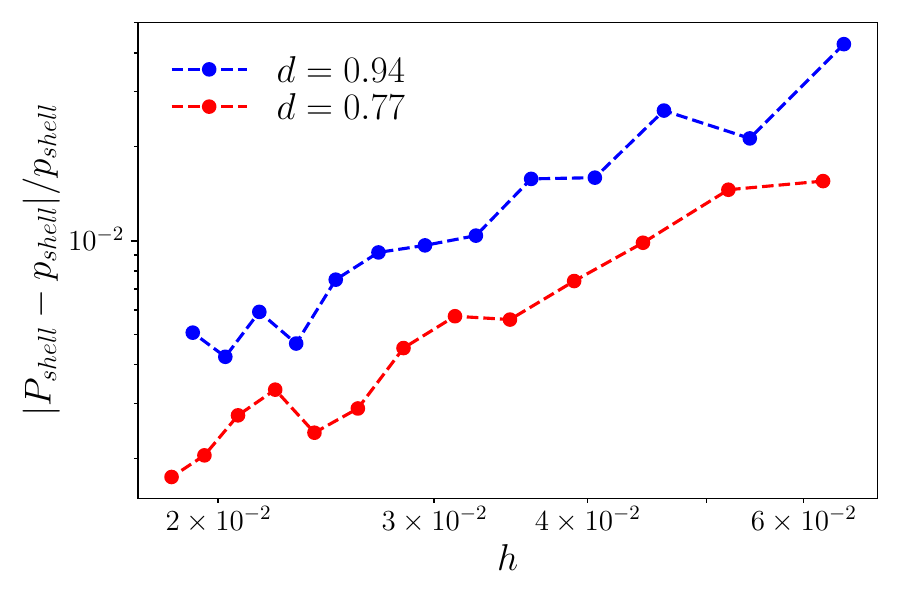}
    \caption{Study of discretization effects. Left panel: The exact continuous probability functional $\continuousp{\mu}{d}$ for the 3-shell where the inner radius is selected as $\innerr{3} = d - \unitr{3}$.  (solid blue line) compared with the corresponding discrete $\discretep{s}{d}$ (black dots) as function of the grasshopper jump distance $d\geq \unitr{3}$. The corresponding probability for the solid 3-ball (red line) is also shown for comparison. For $d>\unitr{3}$ the 3-shell has a higher success probability than the 3-ball.
    Right panel: Relative deviation of $\discretep{s}{d}$ for the 3-shell configuration from the corresponding $\continuousp{\mu}{d}$ as function of the lattice spacing $h$ for two representative values of the grasshopper jump: $d=0.94\approx1.5\unitr{3}$ (blue dots and line) and $d=0.77\approx1.25\unitr{3}$ (red dots and line). The inner radius for each $d$ is $\innerr{3} = d - \unitr{3}$, as before. For the highest resolutions considered ($M\approx 162,000$) the discretization error is below $1\%$. Lines are to guide the eye. \label{fig:shell_probabilities}}
\end{figure}

\begin{figure}
    \centerline{\includegraphics[scale=.8]{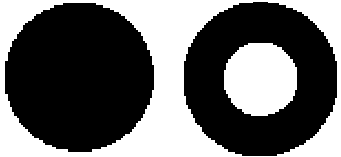}}
    \caption{Left panel: cross section of the optimal configuration for $d=0.64\unitr{3}$ found numerically for a system with $M=160,000$ spins. The configuration has the shape of a solid 3-ball. This configuration was found to be optimal for all $d\leq\unitr{3}$. Right panel: cross section of the optimal configuration for $d=1.32\unitr{3}$ found numerically for a system with $M=160,000$ spins. If the jump length exceeds $\unitr{3}$ the configurations remain isotropic (for $d\lesssim1.4\unitr{3}$) but develop a spherical hole in the center; the radius of the hole grows with increasing $d$. Note that the outer radius of the configuration is slightly larger than $\unitr{3}$ to ensure that it has unit volume.}
    \label{fig:sim_ball}
\end{figure}

In contrast to 2d, in 3d the isotropic solid ball configuration was found to be optimal for jump lengths $d\leq \unitr{3}$, in agreement with our analytical results presented in the preceding sections. For jumps that slightly exceed $\unitr{3}$ the configuration remains isotropic by developing a spherical hole in the center of the ball. We refer to these configurations as shells, specifically 3-shells for $N=3$. It is easy to see that the ball cannot be the optimal solution for $d>\unitr{3}$, since in this case a grasshopper starting out near the center of the ball would land outside of the lawn with probability one. Thus it seems intuitive that an optimal lawn may be obtained by removing the central part with radius of the order of $\innerr{3}=d-\unitr{3}$ from the 3-ball and redistributing it elsewhere. This intuition is confirmed by numerics. In Fig.~\ref{fig:shell_probabilities} we analyze discretization effects for the 3-shell configuration. As for the solid 3-ball, the errors due to discretization are very small. Figure~\ref{fig:sim_ball} shows cross-sections of numerically found optimal grasshopper lawn shapes for two representative jump lengths below and above $\unitr{3}$.

\begin{figure}
\includegraphics[clip=true,width=0.5\textwidth]{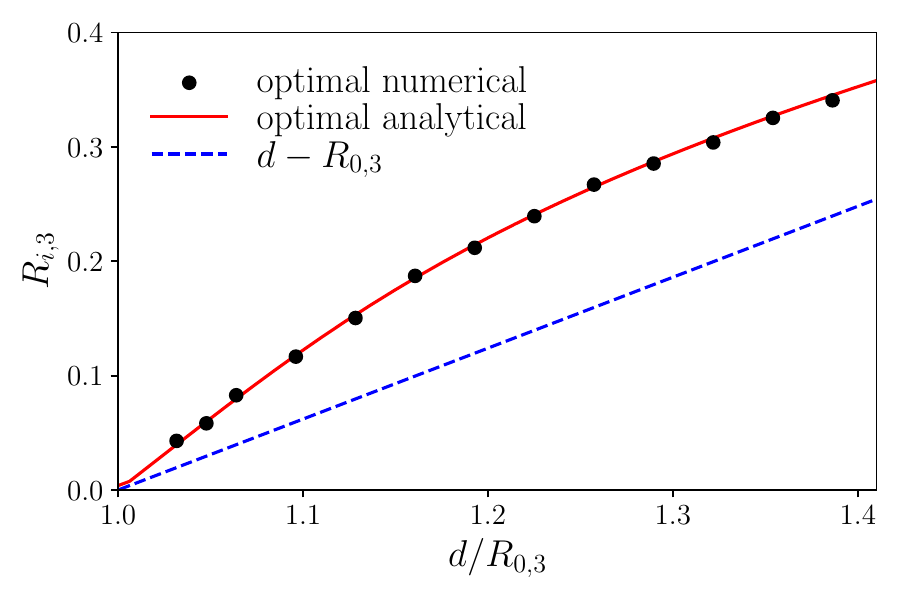}\hfill
\includegraphics[clip=true,width=0.5\textwidth]{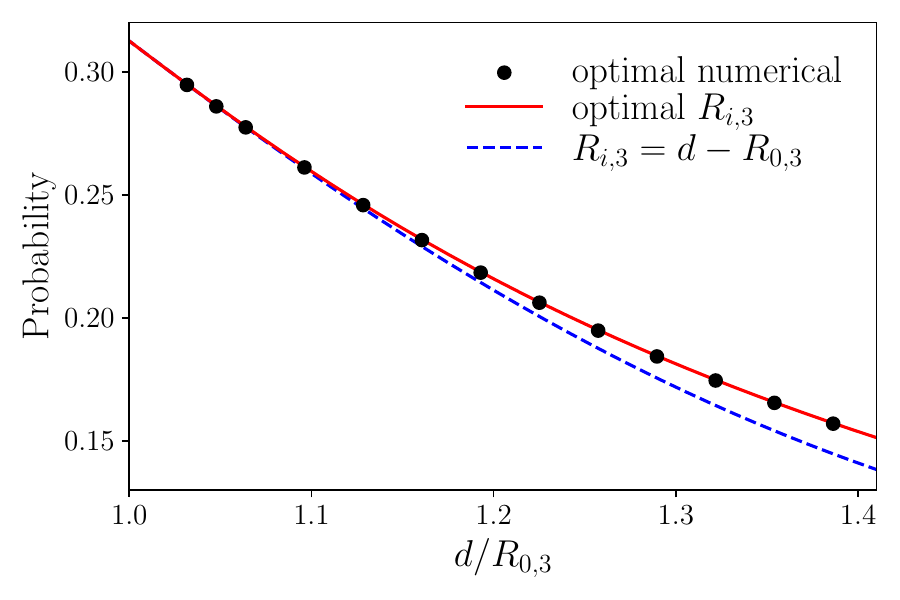}
\caption{Left panel: the optimal inner radius, $\innerr{3}$, of the 3-shell versus grasshopper jump. The numerically found inner radii (black dots) match very well the analytical result (red solid line). The value $d-\unitr{3}$ (blue dashed lines) is shown for comparison. Right panel: the corresponding optimal grasshopper probabilities for the optimal inner radius (red solid line for analytical value and black dots for the numerical value) and for $d-\unitr{3}$ (blue dashed lines). The isotropic 3-shell ceases to be optimal for jumps exceeding a critical value of approximately $1.4\unitr{3}$.}
\label{fig:shell_radii}
\end{figure}
The precise optimal inner radius of the 3-shell configurations can be computed directly, by evaluating the grasshopper probability integral \eqref{successprob} for the radially symmetric 3-shell of a given inner radius in spherical coordinates, and then finding the maximum of the resulting expression as function of inner radius for a given jump. The 3-shell probability at fixed jump length is a simple function of the inner radius with exactly one maximum whose location can be found using standard maximization methods. Fig.~\ref{fig:shell_radii} shows the analytical results together with the numerically found values for different jumps. Numerical simulations show that the isotropic 3-shell configuration ceases to be optimal for jumps exceeding a critical value of approximately $1.4\unitr{3}$, when the shell starts exhibiting periodic perturbations: 8-fold perturbations (resembling a rounded cube) were observed for $1.4\unitr{3}\lesssim d\lesssim 1.58\unitr{3}$ and 6-fold perturbations (resembling a bulbous octahedron) were observed for $1.58\unitr{3}\lesssim d\lesssim 1.68\unitr{3}$. This breaking of full rotational symmetry into a discrete subgroup parallels the symmetry breaking to dihedral symmetry that occurs in the two dimensional model. However, in three dimensions symmetry is only broken for sufficiently large jumps. Both 3d shapes in this category that were observed numerically appear to have full octahedral symmetry (i.e. appear to be invariant under the 48 element group $O_h$)
and appear to correspond to spherical harmonic modulations of the 3-shell. It is conceivable that additional shapes with higher symmetry orders emerge near the transition. These could potentially be observed numerically at higher resolution.

\begin{figure}
    \centering
    \includegraphics[width=0.33\textwidth]{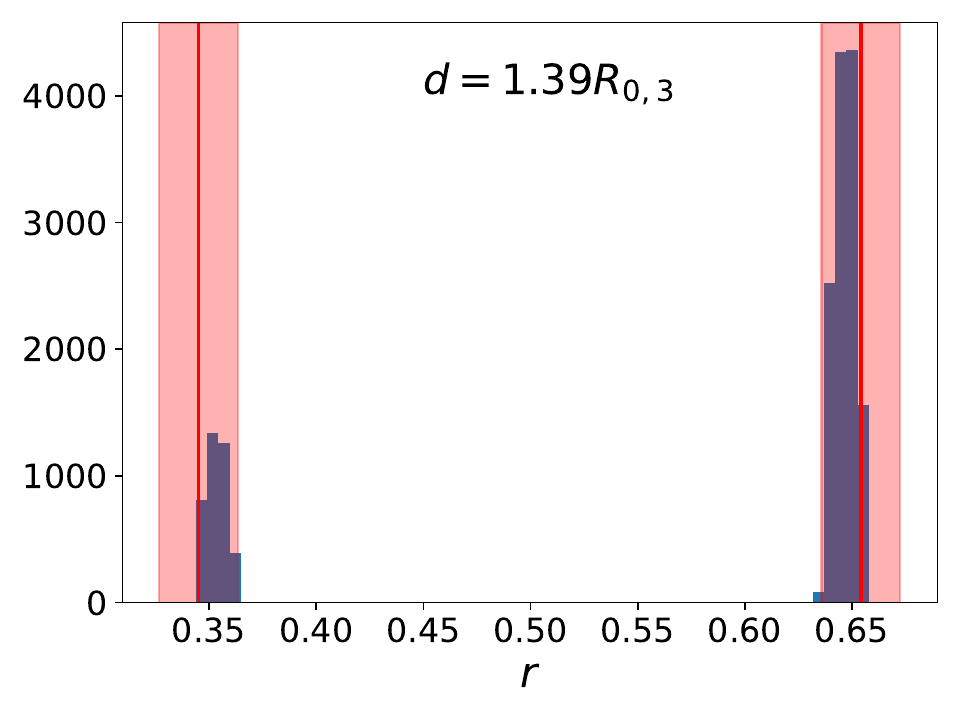}\hfill
    \includegraphics[width=0.33\textwidth]{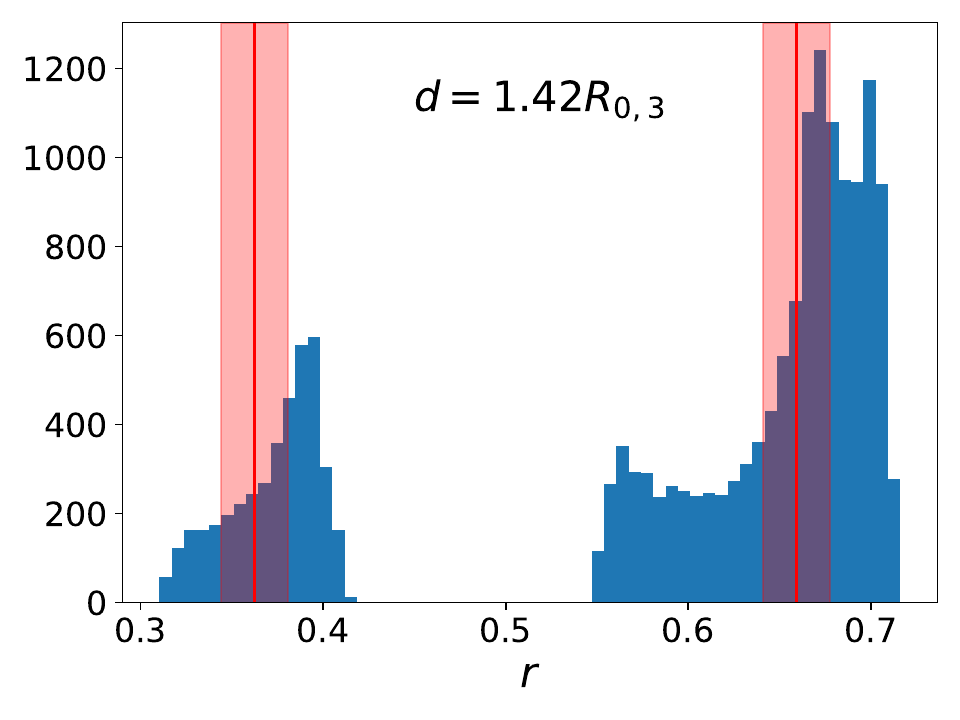}\hfill
    \includegraphics[width=0.33\textwidth]{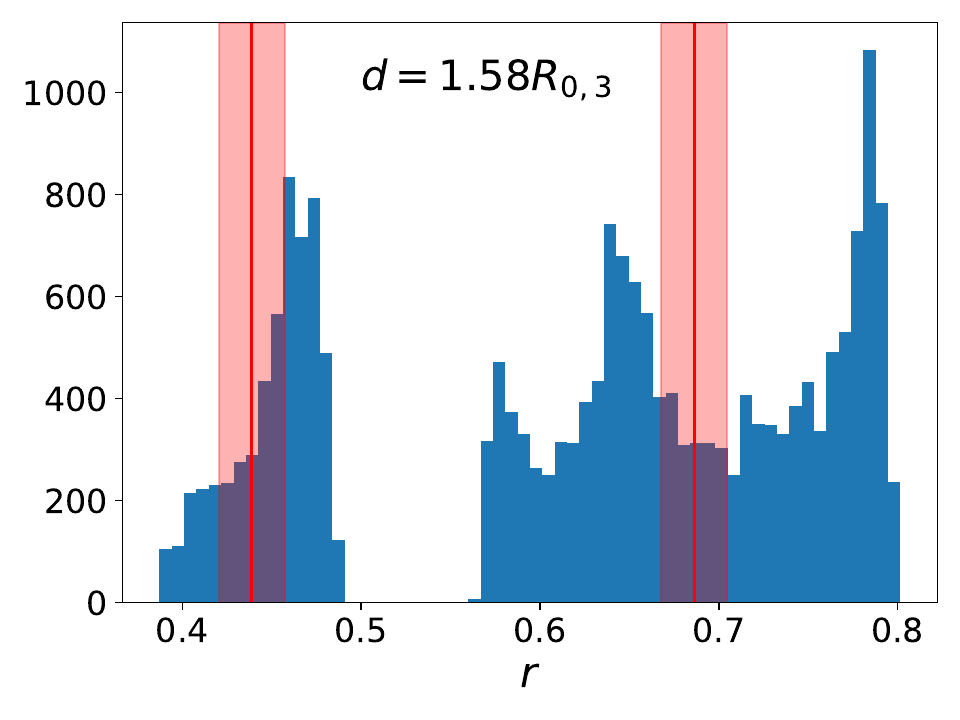}
    \caption{Histograms of the radial coordinates of the configuration boundary points for, from left to right, $d=1.39\unitr{3}$ (isotropic 3-shell), $d=1.42\unitr{3}$ (8-fold perturbation), and $d=1.58\unitr{3}$ (6-fold perturbation). Results shown were obtained for systems with $M=160,000$ spins. The vertical red lines denote the theoretical values for the optimal (for the respective value of $d$) inner and outer radii of the corresponding 3-shell configurations. The shaded regions mark the $\pm h$ interval around the optimal radii.}
    \label{fig:histogram}
\end{figure}
A quantitative procedure to asses whether a given numerical configuration is rotationally symmetric is depicted in Fig.~\ref{fig:histogram}. We histogram the radial coordinates of all boundary points of a configuration, and compare these histograms with the known values of the outer and inner radii of the corresponding optimal isotropic configurations. As can be seen in the left panel of Fig.~\ref{fig:histogram}, in the 3-shell regime the boundary histograms indeed overlap with the expected values, and their width is of the order of the lattice cell width $h$. In the symmetry-broken regimes on the other hand, the distribution widths substantially exceed $h$.

\begin{figure}
\includegraphics[width=\textwidth]{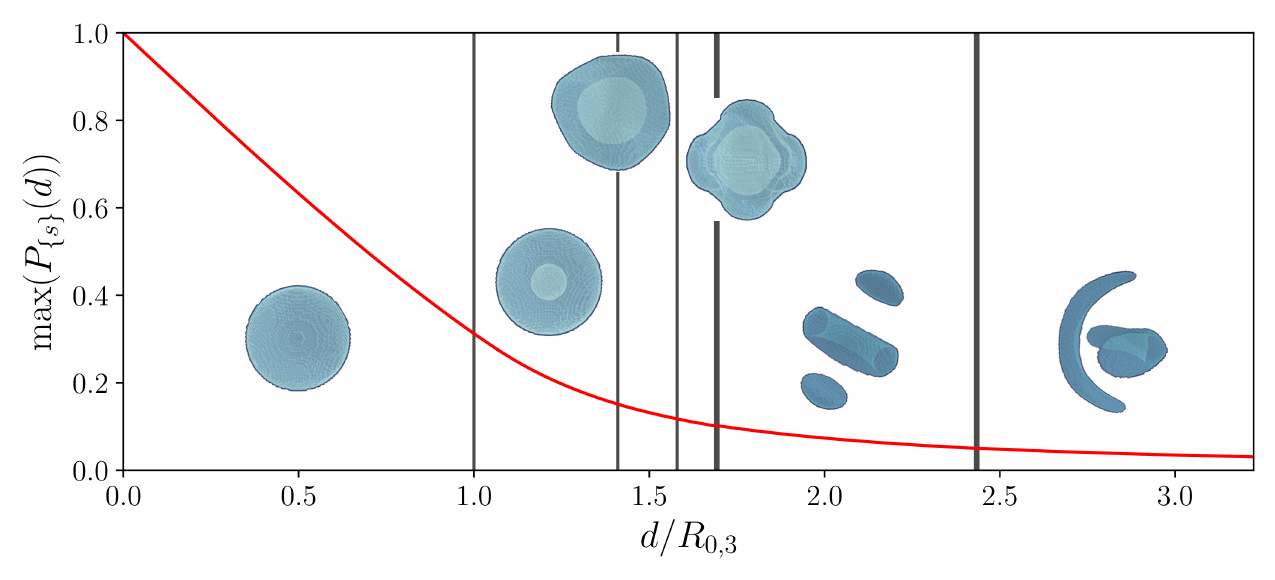}
\caption{Maximal values found numerically for the discrete grasshopper lawn probability $P_{\{s\}}(d)$. Results shown were obtained for systems with $M=40,000$ spins. Vertical lines denote boundaries between the different regimes. From left to right these are: isotropic solid ball, isotropic shells, 8-fold shell perturbations, 6-fold shell perturbations, ring with caps, nested crescents. Insets show examples of representative configurations. 3D animations displaying in full the features summarized here are given in the Supplemental Material.}
\label{fig:numphasediag}
\end{figure}

At even larger jumps the numerically found optimal configurations become disconnected. For $1.7\unitr{3}\lesssim d\lesssim 2.45\unitr{3}$ the optimal shapes consist of a ring with two caps, and for larger jumps of two nested crescents. We show representative shapes obtained numerically and the corresponding phase diagram with optimal probabilities in Fig.~\ref{fig:numphasediag}. For better visualization from different perspectives, animations of the 3d shapes are included in the Supplemental Material.

\section{Summary and discussion}
The setup of the grasshopper problem is rotationally symmetric: the starting point and the direction of the jump are both chosen uniformly. For small jumps, in order for the grasshopper to be able to land outside the lawn, the starting point needs to be close enough to the boundary. Thus one might intuitively assume that the optimal grasshopper lawn shape for small jumps would minimize the boundary. The shape that minimizes the boundary at fixed volume is the $N$-ball, which in two dimensions corresponds to the disk. However, numerical simulations in 2d revealed that optimal lawn shapes were not rotationally symmetric for any jump length, and this result was also proven analytically \cite{goulko2017grasshopper}. For small jumps, full rotational symmetry was found to be reduced to dihedral symmetry $D_n$, where the number $n$ of remaining rotational symmetries is determined by the number of edges of a polygon inscribed in the unit area disk, such that the polygon edge length is close to the length of the grasshopper jump $d$. 

In this work we answered the open question regarding the origin of the symmetry breaking in the grasshopper model. In the limit of small jumps, $d\rightarrow0$, a disk-shaped lawn can be approximated as having zero curvature. We showed that in this minimal model of an infinite half-plane, periodic perturbations of the boundary that are commensurate to the jump lengths $d$ are only marginally stable. For the same model in higher dimensions we showed that all perturbations are stable and the result becomes increasingly insensitive to the commensurability between the perturbation and jump. Numerical simulations in three dimensions confirm that optimal solutions are indeed isotropic: for small jumps the lawns are $3$-balls, which subsequently develop spherical holes as the jump size increases. Rotational symmetry is only broken for large values of the jump ($d\gtrsim1.4\unitr{3}$).

To show that marginally stable perturbations of the minimal model indeed become unstable in the full model, we analyzed the instabilities of the 2d disk with respect to small perturbations of different discrete symmetry breaking orders $n$. We confirmed the dihedral symmetry breaking described in Ref.~\cite{goulko2017grasshopper} and found additional discrete rotationally symmetric solutions that are advantageous compared to the disk. While this perturbative calculation does not directly describe the optimal grasshopper shapes found numerically, there is remarkable agreement between the perturbative predictions and the numerical results. Moreover, the $n\rightarrow\infty$ limit asymptotically approaches the result of the minimal model.

The analytical and numerical techniques we have developed hold promise for broader applications, particularly in elucidating long-range interaction effects in ultracold-atom based quantum emulators and prototype quantum computers. Several optimal grasshopper shapes, including cogwheel patterns, suggest intriguing parallels with dipolar Bose-Einstein condensates, where similar patterns at droplet boundaries have been reported \cite{hertkorn2021dipolar, schmidt2021dipolar}. Another promising application of our techniques is the domain dynamics of quantum Ising spins in Rydberg atom arrays \cite{Browaeys2020rydberg}. Therefore, our work serves not only as an illuminating analysis of symmetry breaking in the grasshopper model but also as a stepping stone for the study of complex quantum many-body systems with long-range interactions. In addition, recent advances in the creation and control of Rydberg atom arrays with  distance-selective interactions \cite{Ebadi2021rydberg, zeiher2022rydberg} present an exciting opportunity to directly build and study two and three dimensional grasshopper-type spin systems in ultracold atoms experiments.

\section*{Acknowledgements}
We are grateful to Ansh Shah for pointing out the connection to dipolar gases. This work is supported by the NSF under Grant No. PHY-2112738 (O.G. and D.L.) and Grant No. PHY-2328774 (O.G.). J.K.-D. is supported by a \textsc{DynSysMath} Specific Initiative of the INFN. A.K. acknowledges financial support from the UK Quantum Communications Hub grant no. EP/T001011/1. 
A.K. was supported in part by Perimeter Institute for Theoretical Physics. Research at Perimeter Institute is supported by the Government of Canada through the Department of Innovation, Science and Economic Development and by the Province of Ontario through the Ministry of Research, Innovation and Science. The Flatiron Institute is a division of the
Simons Foundation.
\bibliographystyle{unsrtnat}
\bibliography{grasshopper}

\appendix

\section{Details of the half-space integration}
\label{sec:half-space.details}

To further evaluate the half-space integral, note that because of the absence of
boundaries, the integration variables can be shifted arbitrarily. We therefore
take $\Delta\mathbf x=\mathbf y-\mathbf x$, yielding
\begin{equation}
  \begin{aligned}
    \delta p_k(d)
    =-\lim_{L\to\infty}\frac1{L^{N-1}}\int_{\mathbb R^{N-1}} d\mathbf x\,d\Delta\mathbf x \, \biggl[&
      k^2\sin(kx_1)\sin\big(k(x_1+\Delta x_1)\big)\,\Phi(\|\Delta\mathbf x\|)+\biggr.\\
      &\biggl.\quad+\frac12\frac{\big[\cos(kx_1)-\cos\big(k(x_1+\Delta x_1)\big)\big]^2}{\|\Delta\mathbf x\|}\Phi'(\|\Delta\mathbf x\|)
    \biggr].
  \end{aligned}
\end{equation}
The integral over $\mathbf x$ can be performed: $N-2$ of the components do not appear in the integrand and therefore produce a factor of $L^{N-2}$, while the component $x_1$ gives the only nontrivial integration. Upon integration and taking the limit, we find
\begin{equation}
    \delta p_k(d)
    =-\frac12\int_{\mathbb R^{N-1}}d\Delta\mathbf x\,\left[
      \frac{\Phi'(\|\Delta\mathbf x\|)}{\|\Delta\mathbf x\|}+\cos(k\Delta x_1)\,\left(k^2\Phi(\|\Delta\mathbf x\|)-\frac{\Phi'(\|\Delta\mathbf x\|)}{\|\Delta\mathbf x\|}\right)
    \right].
\end{equation}
Now we switch to hyperspherical coordinates in $\Delta\mathbf x$, with
\begin{equation} \label{eq:hyperspherical.integral}
  \delta p_k(d)
  =-\frac12\int dr\,r^{N-2}\,d\Omega\,\left[
    \frac{\Phi'(r)}r+\cos(kr\cos\phi_1)\,\left(k^2\Phi(r)-\frac{\Phi'(r)}r\right)
  \right],
\end{equation}
where $d\Omega=\prod_{i=1}^{N-2}\sin^{N-2-i}(\phi_i)\,d\phi_i$. The integrand
is independent of all $\phi$s except $\phi_1$, and the remaining integrals for
$\phi_2,\ldots,\phi_{N-2}$ can be performed to get the volume of an
$(N-3)$-sphere or
\begin{equation}
  \delta p_k(d)
  =-\frac{\pi^\frac{N-2}2}{\Gamma(\frac{N-2}2)}\int dr\,d\phi_1\,\sin^{N-3}(\phi_1)\,r^{N-2}\left[
    \frac{\Phi'(r)}r+\cos(kr\cos\phi_1)\,\left(k^2\Phi(r)-\frac{\Phi'(r)}r\right)
  \right].
\end{equation}
The integral over $\phi_1$ can be evaluated to yield
\begin{equation}
  \delta p_k(d)
  =-\frac{\pi^\frac{N-1}2}{\Gamma(\frac{N-1}2)}\int dr\,r^{N-2}\left[
    \frac{\Phi'(r)}r+{}_0F_1\big(;\tfrac12(N-1);-\tfrac14(kr)^2\big)\,\left(k^2\Phi(r)-\frac{\Phi'(r)}r\right)
  \right],
\end{equation}
where ${}_0F_1$ is a hypergeometric function. Up to here, all this is valid for generic $\Phi$ so long as it vanishes at large argument. Inserting our $\Phi$, the Heaviside function breaks the radial integration into two pieces, giving
\begin{equation}
  \begin{aligned}
    \delta p_k(d)
    &=\frac{\Gamma(\frac N2)}{2\sqrt\pi\Gamma(\frac{N-1}2)}\bigg[
      \frac{k^2}{N-2}\int_0^d dr\,\frac{r^{N-2}}{d^{N-2}}{}_0F_1\big(;\tfrac12(N-1);-\tfrac14(kr)^2\big) \\
    &\qquad  -\int_d^\infty dr\,\left(
      \frac1{r^2}-\left(\frac1{r^2}+\frac{k^2}{N-2}\right){}_0F_1\big(;\tfrac12(N-1);-\tfrac14(kr)^2\big)
    \right)
  \bigg].
  \end{aligned}
\end{equation}
Finally, the integrals over $r$ can be performed, yielding
\begin{equation}
  \begin{aligned}
    \delta p_k(d)
    &=\frac{\Gamma(\frac N2)}{2\sqrt\pi\Gamma(\frac{N-1}2)d}\bigg[
      \frac{(kd)^2}{N-2}\left(\frac{{}_0F_1\big(;\tfrac12(N+1);-\tfrac14(kd)^2\big)}{N-1}
        -{}_1F_2\big(\tfrac12;\tfrac32,\tfrac12(N-1);-\tfrac14(kd)^2\big)
      \right) \\
    &\qquad  +{}_1F_2\big(-\tfrac12;\tfrac12,\tfrac12(N-1);-\tfrac14(kd)^2\big)
    -1
  \bigg],
  \end{aligned}
\end{equation}
which simplifies to
\begin{equation}
  \delta p_k(d)
  =\frac{\Gamma(\frac N2)}{2\sqrt\pi d}\bigg[
    \frac{J_{\frac{N-3}2}(kd)}{(kd/2)^{(N-3)/2}}
    -\frac1{\Gamma(\frac{N-1}2)}
  \bigg],
\end{equation}
as in the main text.
One might notice that in some intermediate steps, the formulae for
$N=2$ are not always well-defined. However, by returning to
\eqref{eq:hyperspherical.integral} and carrying out the calculation for $N=2$,
one finds the same result as that predicted by \eqref{eq:plane.instability}, so this formula is in fact general for $N\geq2$.

\section{Additive perturbations}
\label{sec:additive.perturbations}

Our study of the minimal half-space model examined only perturbations with the
form of a sinusoid along one axis. It is plausible, however, that other forms
of perturbation are more prone to instability. Here, we show that any sum of
finitely many sinusoids, whether along the same or different axes, reduces to
summing the results due to the two sinusoids independently. This strongly
motivates that the instability we examine due to a single sinusoid is the most
relevant for this case, i.e., when the wavelength of the perturbation is very
small compared to the curvature of the surface.

Consider a perturbation that is instead the sum of two plane waves along independent axes, like
\begin{equation}
  h(\mathbf x)=a_1\cos(k_1x_1)+a_2\cos(k_2x_2)
\end{equation}
The first term in \eqref{eq:plane.expansion} explicitly breaks into two independent terms along the two axes. The second is not linear and potentially mixes their contribution. It gives
\begin{equation}
  \begin{aligned}
    \big(h(\mathbf x)-h(\mathbf y)\big)^2
    &=\left[
      a_1\cos(k_1x_1)+a_2\cos(k_2x_2)-a_1\cos(k_1y_1)-a_2\cos(k_2y_2)
    \right]^2 \\
    &=a_1^2\left[\cos(k_1x_1)-\cos(k_1y_1)\right]^2
    +a_2^2\left[\cos(k_2x_2)-\cos(k_2y_2)\right]^2 \\
    &\qquad+2a_1a_2\left[
      \cos(k_1x_1)-\cos(k_1y_1)
    \right]\left[
      \cos(k_2x_2)-\cos(k_2y_2)
    \right]
  \end{aligned}
\end{equation}
The first two terms are again just the sum of the two terms that would result
from individual plane waves, while the third gives a nontrivial mixture. So the
only question to the effect of adding plane waves along different axes is how
this final term contributes to the stability. Changing coordinates to
$\Delta\mathbf x=\mathbf y-\mathbf x$ and integrating in $\mathbf x$ from
$-\frac L2$ and $\frac L2$ (noting that the rest of the term in
\eqref{eq:plane.expansion} is independent of $\mathbf x$ after the coordinate
transformation is made) gives
\begin{equation}
  \begin{aligned}
    &\frac1{L^N}\int_{-\frac L2}^\frac L2d\mathbf x\left[
        \cos(k_1x_1)-\cos\big(k_1(x_1+\Delta x_1)\big)
      \right]\left[
        \cos(k_2x_2)-\cos\big(k_2(x_2+\Delta x_2)\big)
      \right] \\
    \qquad&=\frac1{L^2}\frac8{k_1k_2}\left[
      1-\cos(k_1\Delta x_1)
    \right]\left[
      1-\cos(k_2\Delta x_2)
    \right]\sin(k_1L/2)\sin(k_2L/2)
  \end{aligned}
\end{equation}
Unlike the terms that decompose into additive pieces, this one vanishes in the
limit of large $L$. Therefore, this term contributes nothing to the stability
of the plane.

Now, consider a perturbation that is the sum of two plane waves along the same axes, like
\begin{equation}
  h(\mathbf x)=a_1\cos(k_1x_1)+a_2\cos(k_2x_1)
\end{equation}
with $k_1\neq k_2$. This time, both terms in \eqref{eq:plane.instability} mix the contributions, with the extra expression given by
\begin{equation}
  -a_1a_2\left[
      k_1k_2\left(
        \sin k_1x_1\sin k_2y_1+\sin k_1y_1\sin k_2x_1
      \right)\Phi(\|\mathbf x-\mathbf y\|)
      +\frac12
        (\cos k_1x_1-\cos k_1y_1)(\cos k_2x_1-\cos k_2y_1)
        \frac{\Phi(\|\mathbf x-\mathbf y\|)}{\|\mathbf x-\mathbf y\|}
    \right]
\end{equation}
Like the case of the waves along independent axes above, when we take
$\Delta\mathbf x=\mathbf y-\mathbf x$ and integrate over $\mathbf x$, the
resulting integrals vanish in the limit of large $L$. One gives
\begin{equation}
  \begin{aligned}
    &\frac1{L^N}\int_{-\frac L2}^\frac L2d\mathbf x\,\left(
        \sin k_1x_1\sin k_2(x_1+\Delta x_1)+\sin k_1(x_1+\Delta x_1)\sin k_2x_1
      \right) \\
    &\qquad=\frac2{L^2}\frac{(\cos k_1\Delta x_1+\cos k_2\Delta x_1)}{(k_1-k_2)(k_1+k_2)}
    \left[
      k_2\cos\frac{k_2L}2\sin\frac{k_1L}2-k_1\cos\frac{k_1L}2\sin\frac{k_2L}2
    \right]
  \end{aligned}
\end{equation}
while the other gives
\begin{equation}
  \begin{aligned}
    &\frac1{L^N}\int_{-\frac L2}^\frac L2d\mathbf x\,\left(
      \cos k_1x_1-\cos k_1(x_1+\Delta x_1)\right)\left(\cos k_2x_1-\cos k_2(x_1+\Delta x_1)
      \right) \\
    &\qquad=\frac2{L^2}\frac{1}{(k_1-k_2)(k_1+k_2)}
    \bigg[
      \cos\frac{k_2L}2\sin\frac{k_1L}2\Big(
        k_1(1-\cos k_1\Delta x_1)(1-\cos k_2\Delta x_1)+k_2\sin k_1\Delta x_1\sin k_2\Delta x_1
        \Big) \\
    &\hspace{13em}  -\cos\frac{k_1L}2\sin\frac{k_2L}2\Big(
        k_2(1-\cos k_1\Delta x_1)(1-\cos k_2\Delta x_1)+k_1\sin k_1\Delta x_1\sin k_2\Delta x_1
        \Big)
    \bigg]
  \end{aligned}
\end{equation}

It follows that the stability coefficient resulting from the sum of two plane
waves is given in this case by the sum of the coefficients due to the
individual pure plane waves. Therefore, the effect of more complicated perturbations than those considered here can be reduced to that of the simple ones, with the caveat that these results only strictly hold for perturbations build from sums of finitely many oscillatory terms.

\section{Details of the disk calculation}
\label{sec:disk.details}

\begin{figure}
  \includegraphics{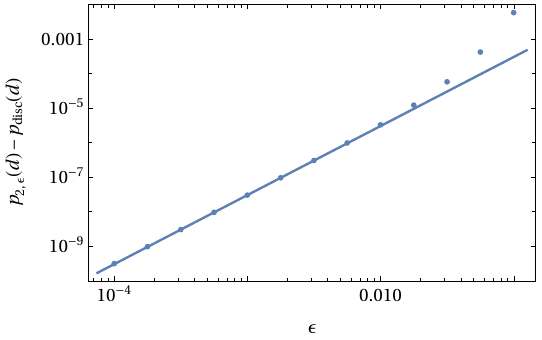}

  \caption{
    The points give the difference in probability between the perturbed disk
    and the disk as a function of perturbation size $\epsilon$ for $n=2$ and
    $d=0.98>d_0$. They were computed using numeric integration on the full
    expression. The solid line gives $\delta p_2(d)\epsilon^2$. The results
    agree at small $\epsilon$, as expected.
  } \label{fig:numeric_check}
\end{figure}

To make the expansion of \eqref{eq:disk.integrals} in $\epsilon$, we need only
expand the integrand, since the limits of integration do not depend on $\epsilon$. The integrand has the form
\begin{equation}
  i(\epsilon)=\Theta\big(g(\epsilon)\big)f(\epsilon)
\end{equation}
where $\Theta$ is the Heaviside theta function and $f$ and $g$ are functions of
$\epsilon$ given by
\begin{equation}
  g(\epsilon)=\|\mathbf r_{n,\epsilon}(\theta)-\mathbf r_{n,\epsilon}(\theta+\phi)\|-d
\end{equation}
and
\begin{equation}
  f(\epsilon)=\frac1{2\pi}\log\left[\frac1d\|\mathbf r_{n,\epsilon}(\theta)-\mathbf r_{n,\epsilon}(\theta+\phi)\|\right]\mathbf r_{n,\epsilon}'(\theta)\cdot\mathbf r_{n,\epsilon}'(\theta+\phi)
\end{equation}
Keep in mind that both $g$ and $f$ are functions of $n$, $\phi$ and
$\theta$, but we have suppressed that dependence for clarity. We need the
second derivative of this expression with respect to $\epsilon$, which gives
\begin{equation}
  i''(0)=
  \Theta\big(g(0)\big)f''(0)
  +\delta\big(g(0)\big)\big[
    2f'(0)g'(0)
    +f(0)g''(0)
  \big]
  +\delta'\big(g(0)\big)f(0)g'(0)^2
\end{equation}
Then the expansion coefficient of the probability for small $\epsilon$ is given by
\begin{equation}
  \delta p_n(d)=-\frac12\int_0^{2\pi}d\theta\int_{-\pi}^\pi d\phi\,i''(0)
\end{equation}
The argument of the Heaviside function $g(\epsilon)$ evaluated at $\epsilon=0$ is independent of $\theta$, and gives
\begin{equation}
  G(\phi)=g(0)=\frac2{\sqrt\pi}|\sin(\phi/2)|-d
\end{equation}
This has two zeros at $\pm\phi_0$ for
\begin{equation}
  \phi_0=\cos^{-1}\left(1-\frac{\pi d^2}2\right)
\end{equation}
and is positive for $|\phi|>\phi_0$. The integrals in $\theta$ can all be
evaluated ignoring the distribution functions, giving
\begin{equation}
  \begin{aligned}
    &i_1(\phi)
    =\int_0^{2\pi}d\theta\,f''(0)
    =-\frac12\log\left(\frac{1-\cos\phi}{1-\cos\phi_0}\right)\left[
        \big((n^2+1)\cos(n\phi)-1\big)\cos\phi-2n\sin\phi\sin(n\phi)
      \right] \\
      &\hspace{5pc}+\frac{\cos\phi\big(1+\cos(n\phi)-2\cos\phi\big)}{2(1-\cos\phi)}
      -4\cos(n\phi/2)^2\sin(\phi/2)^2\frac{\cos\phi-n\sin\phi\tan(n\phi/2)}{1-\cos\phi}
  \end{aligned}
\end{equation}
\begin{equation}
  \begin{aligned}
    &i_2(\phi)=\int_0^{2\pi}d\theta\,\big[
        2f'(0)g'(0)
        +f(0)g''(0)
      \big]
    =\frac1{8\sqrt\pi\,|\sin(\phi/2)|}\bigg\{
      16\cos\phi\cos(n\phi/2)^2\sin(\phi/2)^2
      \\
    &-\left[
        \big((5\cos\phi-3)\cos(n\phi)+\cos\phi-3\big)\cos\phi+8n\sin(\phi/2)^2\sin\phi\sin(n\phi)
        \right]\log\left(\frac{1-\cos\phi}{1-\cos\phi_0}\right)
    \bigg\}
  \end{aligned}
\end{equation}
\begin{equation}
  i_3(\phi)=\int_0^{2\pi}d\theta\,f(0)g'(0)^2
  =\frac1\pi\cos\phi\cos(n\phi/2)^2\sin(\phi/2)^2\log\left(\frac{1-\cos\phi}{1-\cos\phi_0}\right)
\end{equation}
The first piece is proportional to the integral over a Heaviside theta, and
this simply changes the limits of integration. We have
\begin{equation}
  \mathcal I_1=\int_{-\pi}^\pi d\phi\,\Theta\big(G(\phi)\big)i_1(\phi)
  =2\int_{\phi_0}^\pi d\phi\,i_1(\phi)
  =\big[(2-\cos\phi_0)\cos n\phi_0-\cos\phi_0\big]\cot\frac{\phi_0}2
\end{equation}
Next, we have
\begin{equation}
  \mathcal I_2=\int_{-\pi}^\pi d\phi\,\delta\big(G(\phi)\big)i_2(\phi)
  =\int_{-\pi}^\pi d\phi\,\left[
    \frac{\delta(\phi-\phi_0)}{|G'(\phi_0)|}
    +\frac{\delta(\phi+\phi_0)}{|G'(-\phi_0)|}
    \right]i_2(\phi)
    =-4\cos\phi_0\tan\frac{\phi_0}2\cos^2\frac{n\phi_0}2
\end{equation}
Finally, we treat the last term. First, a helpful identity for evaluating an
integral of the derivative of a $\delta$-function convolved with a function.
Note that $\frac\partial{\partial x}\delta(g(x))=\delta'(g(x))g'(x)$. Using
manipulations and integration by parts, we have
\begin{equation}
  \int dx\,\delta'(g(x))f(x)
  =\int dx\,\frac{f(x)}{g'(x)}\frac\partial{\partial x}\delta(g(x))
  =-\int dx\,\delta(g(x))\frac\partial{\partial x}\frac{f(x)}{g'(x)}
  =-\int dx\,\delta(g(x))\left[\frac{f'(x)}{g'(x)}-\frac{f(x)g''(x)}{g'(x)^2}\right]
\end{equation}
which is a general but extremely non-obvious identity. We now can write
\begin{equation}
  \begin{aligned}
    \mathcal I_3
    &=\int_{-\pi}^\pi d\phi\,\delta'(G(\phi))i_3(\phi)
    =-\int_{-\pi}^\pi d\phi\,\left[
      \frac{\delta(\phi-\phi_0)}{|G'(\phi_0)|}
      +\frac{\delta(\phi+\phi_0)}{|G'(-\phi_0)|}
      \right]\left[
        \frac{i_3'(\phi)}{G'(\phi)}-\frac{i_3(\phi)G''(\phi)}{G'(\phi)^2}
      \right] \\
    &=2\cos\phi_0\tan\frac{\phi_0}2\cos^2\frac{n\phi_0}2
  \end{aligned}
\end{equation}
Finally, we have
\begin{equation}
  \delta p_n(d)
  =-\frac12(\mathcal I_1+\mathcal I_2+\mathcal I_3)
  =-\frac{\cos\phi_0-\cos(n\phi_0)}{\sin\phi_0},
\end{equation}
A numerical verification of this result for $n=2$ is shown in Fig.~\ref{fig:numeric_check}.

\end{document}